**The polarization property of nanotubes with taking into account the radial motion of valence electrons**

V.A. Aleksandrov[a], A.V. Stepanov[b], and G.M. Filippov[b]

[a]*Chuvash State University, Russia, Chuvash Republic, Cheboksary, Moskovskiy prospect, 15*

[b]*Cheboksary Politechnic Institute (branch) of the Moscow State Open University, Russia, Chuvash Republic, Cheboksary, P. Lumumba street, 8*

Dielectric properties of carbon nanotubes (CNT) are calculated with taking into account the deviation of carbon valence electrons from to be placed exactly on the CNT surface. The results show difference to the usually applied 2D electron gas theory. The collective as well as individual modes of elementary excitations for a model CNT are presented.

―――――――――――――――――

Corresponding author. Tel.: +7 9656832866, fax: +7 8352 630459

E-mail: filippov38-gm@yandex.ru

# 1. Introduction

The problem of channeling of particles in carbon nanotubes (CNT) attracts a great attention in last two decades (see, e.g., [1-6]) due to intensive developing the researches in nanotechnology. If the channeling particles are charged then extremely important to take into account the polarization field, which each particle created in its vicinity. The knowledge of this field allows the correct estimation of polarization forces as well as the polarization energy losses which could render a great influence on a channeling. For example, in the works [6, 7] authors used these estimations for showing the possibility of rainbow scattering in channeling of protons through the CNT's. The theoretical investigations of the problem are based usually on some simplifications, in particular, on the assumption of 2D electron gas on the CNT surface. It was assumed that the 2D electron gas ensures the same polarization field as it could be for a super-thin metallic nanotube. In this case it might be alleged that for each point charge would be a point image charge, which produced the very high polarization force in the nearest neighborhood of the nanotube's surface. There are the experimental works [8] which can be treated as proved the image existence. However, as it was found, the standard RPA in applying to 2D electronic system don't allow the explanation of the image potential existence, it was used the other model calculations as hydrodynamic [4] or the self-consistent Vlasov approximation [9].

Within the our model of CNT polarization it is assumed that the valence electrons of carbon atoms are found on a background of positive charge of atomic cores which is uniform distributed on a sufficiently thin CNT surface. The positive charge assumed to be distributed with ensuring the translation symmetry along the tube's axis as well as axial symmetry relative to the axis. In transversal relative to the axis direction the positive charge undergo the electrons to attract to the surface. This attraction is described with the potential which has the minimum at the surface. In the vicinity of the surface the potential is assumed parabolic shape. In following we expand any function having the symmetry of the tube's potential in a series $f(\vec{x}) = \sum_{km} e^{ikz+im\varphi} f_{km}(r)$, where $r$ is the radial coordinate in the plane orthogonal to the tube's axis, $k$ – is the projection of the wave vector on an axis, $m$ – is a projection of an angular momentum on a tube's axis (as it is widely accepted, we perform calculations in atomic units unless otherwise stated).

## 2. Eigenfunctions

The eigenfunctions described the motion of valence electrons in the field of the tube's electric charge can be presented in a general form

$$\varphi_{km}(\vec{r}) = C e^{i(kz+m\varphi)} R_{km}(r)$$

where

$$\frac{d^2}{dr^2} R_{km}(r) + \frac{1}{r}\frac{d}{dr} R_{km}(r) + \left(2\mu\varepsilon_\perp - k^2 - \frac{m^2}{r^2} - 2\mu u(r-a)\right) R_{km}(r) = 0. \quad (1)$$

Here $\mu$ - is an effective mass of the electron. The potential may be represented in various forms, e.g.,

$$u(r-a) = \begin{cases} \beta(r-a)^2 - u_0, & |r-a| < \sqrt{u_0/\beta} \\ 0, & |r-a| > \sqrt{u_0/\beta} \end{cases},$$

where $\beta = \mu\omega^2/2$; $\omega \gg 1/\mu a^2$; $\beta \gg 1/\mu a^4$. The last two conditions demand the electron wave function width to be much less compared to the tube's radius. It is more suitable to use a narrow Gauss form of an attractive potential as more proper physically, e.g.,

$$u(r-a) = -u_0 e^{-\beta(r-a)^2/u_0}, \quad \sqrt{u_0/\beta} \ll a$$

Performing the transformation of a function as well as an equation, we get

$$R_{km} = \frac{\chi_{km}}{\sqrt{r}} \; ; \quad \frac{d^2}{dr^2}\chi_{km} + \left(2\mu\varepsilon - k^2 - \frac{m^2 - 1/4}{r^2} - 2\mu u(r-a)\right)\chi_{km} = 0 \;. \qquad (2)$$

Here $\varepsilon$ is the total energy of the electron. Within our approximation when the positive charge is assumed to be distributed homogeneous along the tube, the "radial" energy levels $\varepsilon_{\perp n} = \varepsilon_n - k^2/2\mu - (m^2 - 1/4)/2\mu a^2$ (as well as the radial wave functions $\chi_{mn}$) don't depend on $k$ (we omit subscript $k$ in our notation of this functions) and only weak depend on $m$ for bound states, concentrated in the nearest neighborhood of the point $r = a$.

Within a zero-th order approximation we get

$$\chi_{m0} = \chi_0 = \frac{1}{\pi^{1/4} b^{1/2}} \exp\left(-\frac{(r-a)^2}{2b^2}\right), \qquad (3)$$

where $b = 1/\sqrt{\mu\omega}$ - is an effective width of the electron distribution in the ground state, and

$$\varepsilon_{\perp 0} \approx u(0) + \omega/2; \quad \varepsilon \approx u(0) + \omega/2 + k^2/2\mu + (m^2 - 1/4)/2\mu a^2 .$$

In a general case

$$\varphi_{kmn}(\vec{x}) = \frac{1}{\sqrt{2\pi L}} e^{i(kz+m\varphi)} \frac{\chi_{mn}(r)}{\sqrt{r}} , \quad \int_0^\infty \chi_{mn}^*(r)\chi_{mn'}(r)dr = \delta_{nn'} .$$

## 3. Second quantization

The wave operator of valence electrons

$$\hat{\Psi}(\vec{x}) = \sum_{kmn} \hat{a}_{kmn} \varphi_{kmn}(\vec{x}),$$

where $\hat{a}_{kmn}$ together with conjugated operators $\hat{a}^+_{k'm'n'}$ compose a set of annihilation and creation operators obey the usual set of Fermi commutation relations.

Potential of the electric field created by the nanotube in atomic units obey the equation

$$\Delta \hat{\Phi}(\vec{x}) = -4\pi e (\rho_+(r) - \hat{\rho}_e(\vec{x}))$$

where $e$ - is an elementary charge,

$$\rho_+(\vec{x}) = \begin{cases} 0, & \text{if } |r-a| > \delta r_\perp, \ \delta r \ll a \\ g(r) > 0, & \text{if } |r-a| \le \delta r, \ \delta r \ll a \end{cases}$$

And if $N_a$- number of atoms, $Z_v$ – number valence electrons in each atom, then

$$2\pi L \int_{a-\delta r}^{a+\delta r} g(r) r \, dr = N_a Z_v$$

For each component in the domain $|r-a| > \delta r$ we have

$$\frac{d^2}{dr^2}\hat{\Phi}_{km}(r) + \frac{1}{r}\frac{d}{dr}\hat{\Phi}_{km}(r) - \left(k^2 + \frac{m^2}{r^2}\right)\hat{\Phi}_{km}(r) = 4\pi e \hat{\rho}_{km}(r) \ . \qquad (4)$$

The solution to this equation is

$$\hat{\Phi}_{km}(r,t) = -4\pi e \int_0^\infty I_m(kr_<) K_m(kr_>) \hat{\rho}_{km}(r',t) r' \, dr'; \quad r_< = \min(r,r'), \ r_> = \max(r,r')$$

The Hamiltonian of interaction of tube's electrons takes the form

$$\hat{H}_{ee} = \frac{e^2}{L} \sum_{km} \sum_{\substack{k_1 m_1 \\ n_1 n_2}} \sum_{\substack{k_2 m_2 \\ n_3 n_4}} \hat{a}^+_{k_1,m_1,n_1} \hat{a}_{k_1+k,m_1+m,n_2} \hat{a}^+_{k_2,m_2,n_3} \hat{a}_{k_2-k,m_2-m,n_4} F(k\, m_1 m_2 m, n_1 n_2 n_3 n_4) \,, \qquad (5)$$

where

$$F(k\, m_1 m_2 m, n_1 n_2 n_3 n_4) = \int_0^\infty dr \chi^*_{m_1 n_1}(r) \left( \int_0^\infty dr' I_m(kr_<) K_m(kr_>) \chi^*_{m_2 n_3}(r') \chi_{m_2-m,n_4}(r') \right) \chi_{m_1+m,n_2}(r)$$

In the approach, in which the radial movement does not change, numbers $n_i$ will coincide among themselves. The integral here only weakly depends on $k_{1,2}, m_{1,2}$ Taking into account an independence of "transverse" wave functions $\chi_{mn}$ on the wave number $k$, and assuming the main quantum number $n = 0$ don't change in collisions, and $\chi_{m0}(r) \approx \chi_0(r)$, we obtain $F(k\, m_1 m_2 m, n_1 n_2 n_3 n_4) \approx F(k,m)$, where

$$F(k,m) = \int_0^\infty dr \chi^*_0(r) \left( \int_0^\infty dr' I_m(kr_<) K_m(kr_>) \chi^*_0(r') \chi_0(r') \right) \chi_0(r) \,. \qquad (6)$$

In the simplest approximation we get $F(k,m) \approx I_m(ka) K_m(ka)$.

The difference between $I_m(ka) K_m(ka)$ and $F(k,m)$ is illustrated on the Fig. 1. On this figure the dependence of the product $I_m(ka) K_m(ka)$ and the function $F(k,m)$ on the variable $ka$ is compared for the wave function width $b = a/10$. We see the radial motion of tube's electrons can significantly diminish the electron-electron interaction in nanotube.

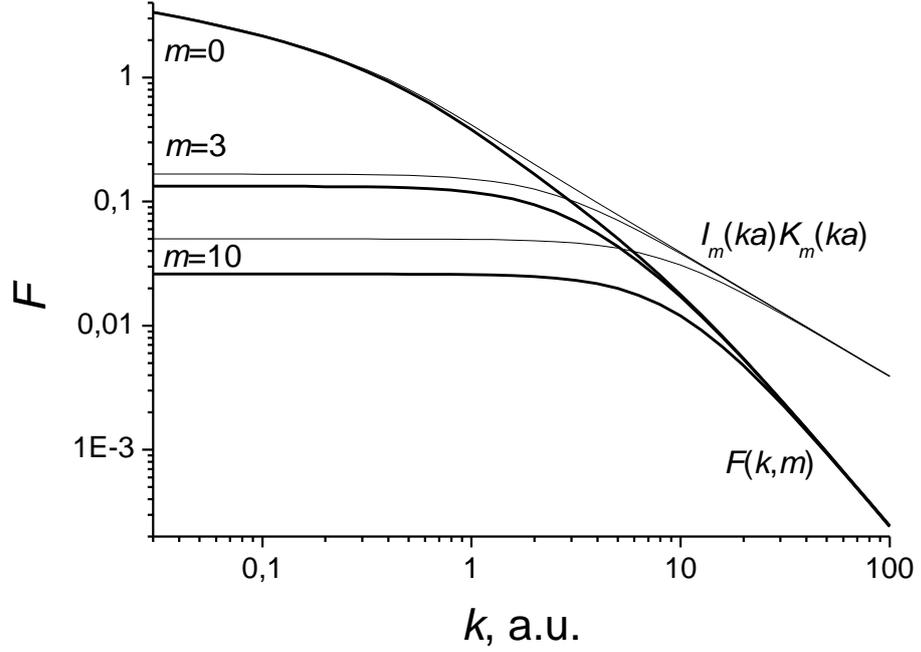

Fig 1. Thin lines: $I_m(ka)K_m(ka)$; thick lines: $F(k,m)$ as functions on $ka$. The curves are drawing for $b = a/10$: Comparison of the 2D electron gas approximation of the function $F(k,m)$ and a new one.

## 4. Dielectric function

In the Fig. 2 a diagram of scattering of two charges (two external electrons) found in CNT is presented. The ring of dash lines relates to the product of propagators of electrons from CNT valence electron system, which should be integrated as on the energy as well on the momentum, which are not pre-defined by the conservation laws. In order to consider only comparatively small deviations from the 2D electron gas approximation, in following we restrict ourselves with transitions only at $n=n'=0$. In this case, performing the calculation within the usual Random Phase Approximation (RPA) procedure, with summing all ring diagrams, we obtain the dielectric function

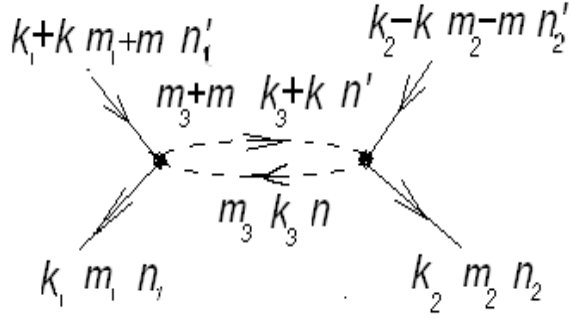

Fig. 2.

$$\varepsilon_{km}(\omega) \approx 1 + 2\frac{e^2}{L} F(k,m) \sum_{k_1 m_1} \frac{n_{k_1,m_1} - n_{k_1+k,m_1+m}}{\omega - \varepsilon_{k_1+k,m_1+m} + \varepsilon_{k_1,m_1} + i(\delta_{k_1+k,m_1+m} - \delta_{k_1,m_1})} \ , \qquad (7)$$

where

$$\varepsilon_{k,m} = u(0) + \omega/2 + k^2/2\mu + (m^2 - 1/4)/2\mu a^2 \ ; \quad \delta_{k,m} = +0 \cdot sign(\varepsilon_{k,m} - \varepsilon_F); \quad k_F = \sqrt{2\mu\varepsilon_F} \ .$$

Generally speaking, in the last formula it doesn't correct to replace the sum over $k_1$ with the corresponding integral. It is because $k_1$ changes not continuously but discrete. Nevertheless, in the area found out of the poles this transformation is quiet correct and gives the small deviation of the order $\sim (k_F L)^{-1}$ from the exact result. But in the nearest neighborhood of each pole $\omega \approx \pm \omega_{kk_1 mm_1}$, where

$$\omega_{kk_1 mm_1} = \left[ \frac{k}{\mu}\left(k_1 + \frac{k}{2}\right) + \frac{m}{\mu a^2}\left(m_1 + \frac{m}{2}\right) \right], \quad -\sqrt{k_F^2 - \frac{m_1^2}{a^2}} < k_1 < \sqrt{k_F^2 - \frac{m_1^2}{a^2}}$$

this representation isn't correct. In this case we can take into account exactly the main polar term and some terms in the sum nearest to the main. For the other part of the sum it is sufficient to apply the integral approximation. In this approach we totally exclude the imaginary parts of generalized functions taking part in the sum over $k_1$. The dielectric function becomes real and contains only main parts of logarithms. After performing the integration one can get

$$\varepsilon_{km}(\omega) = 1 + 2\frac{e^2}{L}F(k,m)\sum_{k_1 m_1}\left(\frac{1}{\omega - \omega_{kk_1 mm_1} + i0} - \frac{1}{\omega + \omega_{kk_1 mm_1} - i0}\right) \approx$$

$$1 + \frac{\mu e^2}{\pi k}F(k,m)\sum_{m_1}\ln\left|\frac{(\sqrt{k_F^2 - m_1^2/a^2} + \tilde{k})^2 - (\mu\omega/k)^2}{(\sqrt{k_F^2 - m_1^2/a^2} - \tilde{k})^2 - (\mu\omega/k)^2}\right| + \Delta\varepsilon_{km}(\omega)$$

where the last term contains some near the particular pole exactly accountable terms. It is to add the factor 2 due to the spin degree of freedom. In result we have:

$$\varepsilon_{km}(\omega) \approx 1 + 2\frac{\mu e^2}{\pi k}F(k,m)\sum_{m_1}\ln\left|\frac{\omega^2 - [(k/\mu)\sqrt{k_F^2 - m_1^2/a^2} + k^2/2\mu + m(m_1 + m/2)/\mu a^2]^2}{\omega^2 - [(k/\mu)\sqrt{k_F^2 - m_1^2/a^2} - k^2/2\mu - m(m_1 + m/2)/\mu a^2]^2}\right| + \Delta\varepsilon_{km}(\omega)$$

here $-(m_1)_{max} \leq m_1 \leq (m_1)_{max}$, $(m_1)_{max} = [ak_F]$. In the case $F(k,m) \to I_m(ka)K_m(ka)$ we obtain the result which first published in the works [1, 2].

## 5. Elementary excitations

It is well known that elementary excitations of polarization media arisen as the solutions to the equation $\varepsilon(k,\omega) = 0$ (see, e.g., [10]). In the case of nanotube we get the similar equation for each set ($k, m$), id. est.

$$\varepsilon_{km}(\omega) = 0 .$$

As it is usual in solid state problem, in our case we have the collective mode which take place in the region out of $\varepsilon_{km}(\omega)$ poles, and individual modes taken place in the neighborhood of poles. In a fig.3 the sample of elementary excitations for a nanotube of arm-chair type (5,5) is depicted. The radius of a tube approximately coincides to the radius of fullerene $C_{60}$. An our approximation is used, a quantity $F(k,m)$ is calculated in our model with $b = a/10$. It is assumed 0.2 electrons on each carbon atom included in the electron gas. In this case $m_{1\max} = 2, k_F = 0.3667$. Thick lines correspond to the collective excitations as solutions to the equation $\varepsilon_{km}(\omega) = 0$. Thin solid lines correspond to bounds of individual excitation bands

$$\omega_{\pm}(k) = \pm k\sqrt{k_F^2 - (m_1/a)^2} + k^2/2 + m(m_1 + m/2)/a^2.$$

We see the characteristic "moustaches" of collective modes, which can be considered as the sources of boundary lines of individual excitation bands. There are many individual excitation bands (in the fig.3 only two is shown) which intersect as a main collective excitation branch as one another. The other collective branches are found at $m \neq 0$. They don't turn to zero at $k \to 0$ in opposite to the main case $m = 0$. For the main collective band we found a like-acoustic behavior of excitations at $k \to 0$ which could not clarify the anticipated metal-like polarization properties of a carbon nanotube. For $m \neq 0$ this explanation is probable but obviously not in the same manner as it would be possible for a real metal.

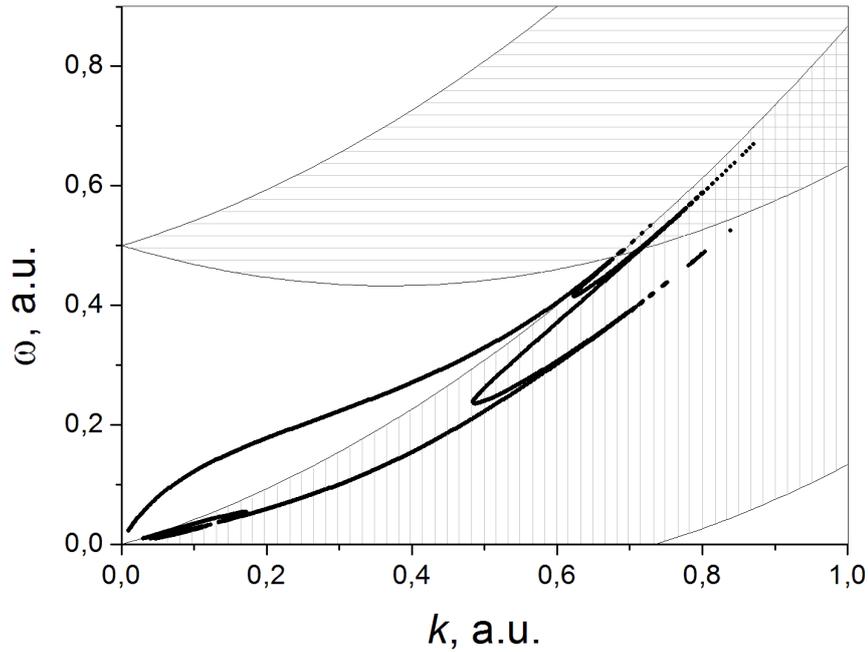

Fig. 3. Zeros of dielectric function $\varepsilon_{km}(\omega)$. For $F(k,m)$ is used the simplest approximation, $I_m(ka)K_m(ka)$. Arm-chair tube (5,5). Elementary excitations of nanotube are shown at $k_F$ =0.3667 a.u., $a$=6.6 a.u. Thick lines are for collective modes, where $\mathrm{Re}[\varepsilon_{km}(\omega)]=0$. Thin lines describe the boundaries of individual mode bands. On the picture only two bands are shown - for $m=0,1$ and $m_1=0$, which intersect the main collective mode branch.

## 4. Conclusions

Taking into account the possibility of radial motion of valence electrons in carbon atoms leads to diminishing the influence of electron-electron interaction on the polarization properties of CNT compared to the 2D electron gas approximation. Elementary excitations of CNT electronic system contain as well as collective as individual modes. The points corresponded to collective excitations are placed on several almost closed curves having specific "moustaches" on a $(k,\omega)$- plane. The dispersion curves for individual modes represent a set of parabolas on a $(k,\omega)$- plane.